\documentclass[fleqn,10pt]{wlscirep}
\usepackage[utf8]{inputenc}
\usepackage[T1]{fontenc}
\usepackage{xcolor}
\title{Secondary electron emission under magnetic constraint: from Monte Carlo simulations to analytical solution}

\author[]{Claudiu Costin}
\affil[]{Iasi Plasma Advanced Research Center (IPARC), Faculty of Physics, Alexandru Ioan Cuza University of Iasi, Iasi-700506, Romania}

\affil[]{claudiu.costin@uaic.ro}

\begin{abstract}
The secondary electron emission process is essential for the optimal operation of a wide range of applications, including fusion reactors, high-energy accelerators, or spacecraft. The process can be influenced and controlled by the use of a magnetic field. An analytical solution is proposed to describe the secondary electron emission process in an oblique magnetic field. It was derived from Monte Carlo simulations. The analytical formula captures the influence of the magnetic field magnitude and tilt, electron emission energy, electron reflection on the surface, and electric field intensity on the secondary emission process. The last two parameters increase the effective emission while the others act the opposite. The electric field effect is equivalent to a reduction of the magnetic field tilt. A very good agreement is shown between the analytical and numerical results for a wide range of parameters. The analytical solution is a convenient tool for the theoretical study and design of magnetically assisted applications, providing realistic input for subsequent simulations.
\end{abstract}
\begin{document}

\flushbottom
\maketitle
%
%
\thispagestyle{empty}

\section*{Introduction}

The International Thermonuclear Experimental Reactor (ITER), the Large Hadron Collider (LHC), and the International Space Station (ISS) are the most ambitious scientific research experiments in their fields, respectively the nuclear fusion research, high-energy particle physics, and space science. Involving huge human and financial resources and worldwide collaborations, they have different goals but they must overcome similar issues. A common issue for the three experiments is the secondary electron emission (SEE) process, which is continuously investigated as part of the plasma-surface interaction. The SEE process consists of releasing electrons from solid surfaces under the bombardment of primary particles: neutrals, charged particles, or photons \cite{Chap80}. Quantitatively, the process is characterized by the SEE yield (SEY), defined as the number of secondary electrons released by a single primary particle. The symbol of the SEY is $\delta$ when primary particles are electrons and $\gamma$ for the others. 

It is well known that the SEY depends on both the primary particle and surface. With respect to the primary particle, the SEY depends on the particle nature (electron, ion, neutral, or photon), mass, electric charge state, energy, and angle of incidence. Regarding the surface, the SEY is influenced by the surface material, crystallographic structure, temperature, cleanliness condition, etc. Consistent reviews provide comprehensive reference lists for both electron \cite{Brui54} and ion \cite{Lay86} induced SEE.

Regardless of the nature of primary particles, secondary electrons are an important component of the plasma-surface interface. They contribute to the balance of charged particles to the surface and space charge formation. In this regard, various concerns are associated with the SEE process. The secondary electrons cool the plasma and lower the potential drop over the sheath, increasing the heat flux and the electron losses to the surface \cite{Hobb67}. For a SEY larger than 1, the space charge may completely vanish \cite{Camp12}, leaving the surface fully exposed to plasma. Such effects are of major importance for the optimal operation of fusion devices \cite{Harb78} since SEYs close to 1 have been reported in the near-wall region of a tokamak \cite{Gunn12}. The same effects and the change of the near-wall conductivity also affect the operation of space propulsion Hall thrusters \cite{Moro00,Rait11}. Secondary electrons are responsible for the generation of instabilities in both high-energy accelerators \cite{Izaw95,Zimm04} and low temperature plasma \cite{Hend94,Gris99}. The performance of high-energy accelerators is limited by the electron cloud effect as a result of the secondary emission \cite{Zimm04,Cimi04}. In space applications, the SEE process leads to electrostatic charging of spacecraft and satellites, affecting on-board electronic devices \cite{Lai02}. Therefore, for such applications, it is crucial to assure a low SEY.

On the other hand, the SEE process has its benefits. For all electrical discharges governed by Townsend’s breakdown theory, the process of SEE induced by ion bombardment at the cathode is essential for both ignition and maintaining the discharge \cite{Chap80,Bara11}. Secondary electrons are used in scanning electron microscopy to form images \cite{Seil83} or in photomultiplier devices to amplify signals \cite{Tao16}. Also, the SEY can be used as an indicator for the surface cleanliness \cite{Gonz17}. 

Whether it is treated as an issue or not, the control of the SEE process and the accurate knowledge of the SEY are of great importance for both applications and numerical simulations. Surface morphology has been shown to play an important role in the emission process. Natural or induced surface roughness may diminish or increase the SEY. For example, fuzz surfaces obtained by exposing W targets to He$^{+}$ ion bombardment can reduce the SEY by more than 50\% \cite{Pati16}. Such effect is expected to influence the operation of ITER, in which W is the leading candidate material for the divertor region and He will result from the fusion reaction \cite{Pitt13}. The roughness of a fuzz surface is not manageable by the user when it is produced inside a fusion reactor. In contrast, a featured surface allows controlling the SEY by adjusting the shape and the aspect ratio of the pattern \cite{Pivi08,Baje20,Agui13}. The pattern can be regular, as in the case of continuously shaped surfaces with different groove profiles \cite{Pivi08,Baje20}, or irregular, as in the case of velvet surfaces (lattices of normally-oriented fibers) \cite{Agui13}. 

The SEE process can also be influenced and controlled by the presence of a magnetic field to the emitting surface. By changing the electronic and magnetic state of materials \cite{Masr16,Apet10}, the magnetic field affects the internal mechanism of electron emission and, consequently, modifies the SEY. Moreover, the magnetic field has a major effect on the electrons that already escaped from the surface. The magnetic field guides the secondary electrons on helical trajectories. It forces a certain number of electrons to return to the emitting surface where they might be reflected or recaptured. If recaptured, the electrons will not generate any relevant effect in plasma/vacuum and they should not be considered for further calculations. The direct consequence of electron recapture is a decrease in the number of secondary electrons able to contribute to the evolution of the investigated system. Therefore, the SEY itself is no longer significant for such a case and it should be replaced by a new parameter which is often referred to as the effective SEY, $\delta_{eff}$ or $\gamma_{eff}$. 

The effective SEY is obtained by subtracting the number of recaptured electrons from the number of secondary electrons released by a single primary particle. It is a crucial parameter in various applications assisted by a magnetic field, such as: magnetically confined plasma in fusion devices \cite{Taka96}, Hall thrusters \cite{Moro00,Rait11}, hollow cathode discharges \cite{Tiro18,Bhuv19}, or magnetron sputtering reactors \cite{Buyl04,Cost05,Bara11}; electron guidance by magnetic immersion lenses in scanning electron microscopy \cite{Toth06}; magnetic suppression of the SEE from the beam screen of a high-energy accelerator \cite{Anas01} or from the negative electrode of a beam direct energy converter \cite{Hash90}. Therefore, the effective secondary electron emission under magnetic field influence has been investigated, independently or in connection with the mentioned applications.

The change of the SEY due to a magnetic field has been measured \cite{Taka96,Anas01,Hash90,Tan06,Pivi08} and calculated, both analytically \cite{Sato94,Cost05,Tskh00,Igit01} and numerically \cite{Pivi08,Buyl04,Mizo95,Nish96,Tskh00,Igit01}. Calculation results are in line with the experimental findings. To ensure a generally valid description of the SEE process under magnetic field influence, the physical quantity which is often calculated by numerical simulations is the relative SEY, $f$. It is defined as the effective SEY normalized to the SEY, $f=\delta_{eff}/\delta$ or $f=\gamma_{eff}/\gamma$. The relative SEY does not describe the internal mechanism of electron emission. It is a fraction of the total number of secondary electrons emitted by the surface (ranging between 0 and 1), which contains only the electrons that are relevant for the investigated system. The relative SEY is independent of all parameters that influence the SEY. However, it depends on the magnetic field magnitude and tilt, the velocity (speed and orientation) distribution function of the emitted electrons, surface recapture probability, surface morphology, and the presence of an electric field to the surface.

For open magnetic field lines, the influence of surface morphology on the SEY has been investigated in \cite{Pivi08} with respect to the energy of incident particles. Self-consistent analysis of the secondary electron emission process and sheath formation is reported in \cite{Igit01,Sato94}. In the self-consistent approach, the electric field in the sheath and the effective SEY are dependent on each other. The dependence of the relative SEY on the energy of the secondary electrons has been analyzed in \cite{Tskh00}. A detailed analysis of the relative SEY is reported in \cite{Mizo95}, considering most of the depending parameters (electron reflection on the surface is not included). Analytical investigations of the effective SEY \cite{Igit01,Tskh00,Sato94} resulted in integral based formulas, which have to be estimated for each specific condition. For the particular case of a magnetron discharge, a detailed numerical analysis is reported in \cite{Buyl04}. Also, a simple formula of the relative SEY has been derived from a fluid model \cite{Cost05}, but with only a few parameters and truncated reflection. However, a consistent analytical solution valid for a wide range of parameters, summing up the previous findings and offering an easy-to-use tool is still required.

This study uses a three-dimensional Monte Carlo (MC) simulation method to investigate the SEE process in an oblique magnetic field $B$. It explores the effects produced on the relative SEY $f$ by the magnetic field magnitude and tilt, electron reflection, electron emission energy, and electric field in front of the surface. The role of the angular distribution of the secondary electrons is also investigated. An analytical formula is derived for the relative SEY, based on the analysis of the numerical results. 
It is specific for a cosine angular distribution of the secondary electrons. With respect to the other parameters, the analytical formula has a wide range of applications. Simulations are made for a flat surface. The electrons trapped by the asperities of a rough surface, due to the presence of the magnetic field \cite{Pivi08}, are not considered to escape from the surface. In this regard, the surface roughness affects the SEY and not the relative SEY. The effect of the electric field on the electron binding energy, which generates the so-called field emission process, also affects the SEY and not the relative SEY. Consequently, the latter two phenomena are not captured by the analytical formula. Because the relative SEY does not describe the internal mechanism of electron emission, the proposed analytical formula also applies to secondary electrons emitted in transmission configuration. 

\section*{Monte Carlo method}

In the MC simulation, $N_{0}$ secondary electrons are randomly released from the surface obeying a cosine angular distribution, which is generally accepted to define the secondary electron emission process, regardless of the nature of the primary particle \cite{Brui54,Kawa99}. A correct angular distribution has to be defined with respect to the solid angle, resulting in the following formula for the cosine distribution \cite{Cost20}:  
\begin{equation}
g_\Omega(\theta)=\frac{dN(\theta)}{N_0d\Omega}=\frac{1}{\pi}\cos\theta
\label{eq:1},
\end{equation}
 with $dN(\theta)$ being the number of electrons emitted in the solid angle $d\Omega=\sin\theta d\theta d\varphi$ defined by the polar angle $\theta$ and the azimuthal angle $\varphi$. The polar angle $\theta$ is measured with respect to the surface normal, varying from zero to $\pi/2$, since electrons are emitted on a single side of a planar surface. To assure a cosine angular distribution, the angle $\theta$ is generated as \cite{Cost20}:
\begin{equation}
  \theta = \arccos{\sqrt{1-r}},
  \label{eq:2}
\end{equation}
with $r$ being a random number between 0 and 1. The azimuthal angle $\varphi=2\pi r'$ is uniformly generated between zero and $2\pi$, using a new random number $r'$ ranging between 0 and 1. 

The energy $\epsilon$ of the secondary electrons is randomly sampled according to a Maxwell-Boltzmann like distribution:
\begin{equation}
f_{MB}(\epsilon)=\frac{dN(\epsilon)}{N_0d\epsilon}=C\sqrt\epsilon\exp(-\frac{\epsilon}{\epsilon_S})
\label{eq:3},
\end{equation}
using the acceptance/rejection method \cite{Gentle98}. In eq. (\ref{eq:3}),  $C$ is a normalization constant and $\epsilon_S$ is the energy corresponding to the most probable speed of the secondary electrons. According to \cite{Ster57} the energy distribution function (EDF) of the secondary electrons has basically the same form for all metals, independent of the work function. Unlike other distributions in the literature \cite{Chun74,Bouc80,Scho96,Caza12}, eq. (\ref{eq:3}) is independent of the surface material and has the advantage of a single fitting parameter. It can also be seen as a simplified form of the EDF derived in \cite{Bouc80}. With the right choice of $\epsilon_S$, usually below 10-15 eV \cite{Chap80}, eq. (\ref{eq:3}) is a good approximation of different secondary electron distributions reported in the literature \cite{Chap80,Izaw95,Brui54,Kawa99,Pati15}.

The secondary electrons are moving in a low background pressure, on collisionless trajectories (the collision frequency is much lower than the cyclotron frequency). Each trajectory is integrated using the leap-frog algorithm coupled with the Boris scheme, since it is known to achieve a good balance between accuracy, efficiency and stability for an imposed time step limit $\omega_{c}\Delta t\leq 0.2$, where $\omega_{c}=\frac{eB}{m_e}$ is the electron cyclotron frequency \cite{Bird91}. Secondary electrons that return to the surface can be either reflected or recaptured, a process described by the reflection coefficient $R$. If reflected, the electron is returned in the simulation space having the same speed as the incident one, angularly distributed according to (\ref{eq:1}). Some of the electrons may experience multiple reflections. 

In the present simulation, the time step is $1\%$ of the electron cyclotron period, which corresponds to $\omega_{c}\Delta t\approx 0.06$. Each electron is tracked either until it is recaptured by the surface or a total integration time of 20 electron cyclotron periods. The latter allows treating a large number of successive reflections, assuring the convergence of eq. (\ref{eq:5}) to eq. (\ref{eq:6}) for all investigated conditions. $N_0$ is $10^6$ for each computation.

\begin{figure}
\includegraphics[width=8.6cm]{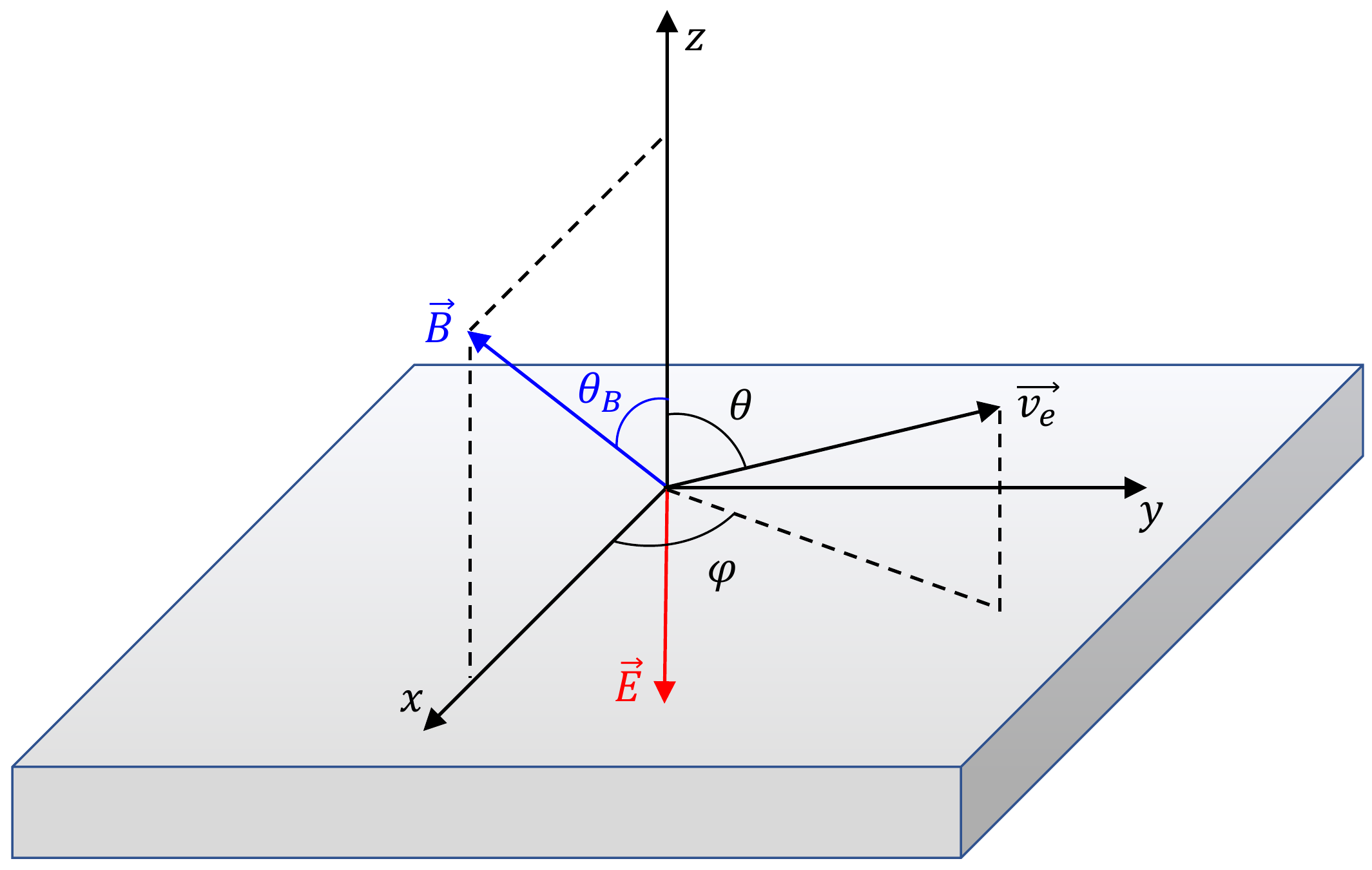}
\centering
\caption{Schematic representation of magnetic field $B$, electric field $E$ and secondary electron velocity $v_e$ vectors.}
\label{fig:Fig0}
\end{figure}

The magnetic field is homogeneous, tilted by a polar angle $\theta_B$ relative to the surface normal. Magnetic field lines are open, leaving the surface and closing to infinity. An electrostatic sheath is considered in front of the surface, with a constant electric field $E$ pointing perpendicularly towards the surface. The electric field repels the secondary electrons from the surface. It acts along the entire trajectory of the secondary electrons, assuming that the sheath thickness is twice as large as the Larmor radius of the secondary electrons. This is a valid assumption for magnetic fields of the order of $0.1-1\text{ T}$, as in magnetron sputtering devices \cite{Buyl04} and tokamaks \cite{Mizo95}, but it may fail for lower magnetic fields ($\sim0.01\text{ T}$), as in Hall thrusters \cite{Rait11} or hollow cathode discharges \cite{Tiro18,Bhuv19}. The schematic representation of magnetic field $B$, electric field $E$ and secondary electron velocity $v_e$ vectors is plotted in Fig. \ref{fig:Fig0}.

\section*{Analytical solution and results}

\begin{figure}
\includegraphics[width=8.6cm]{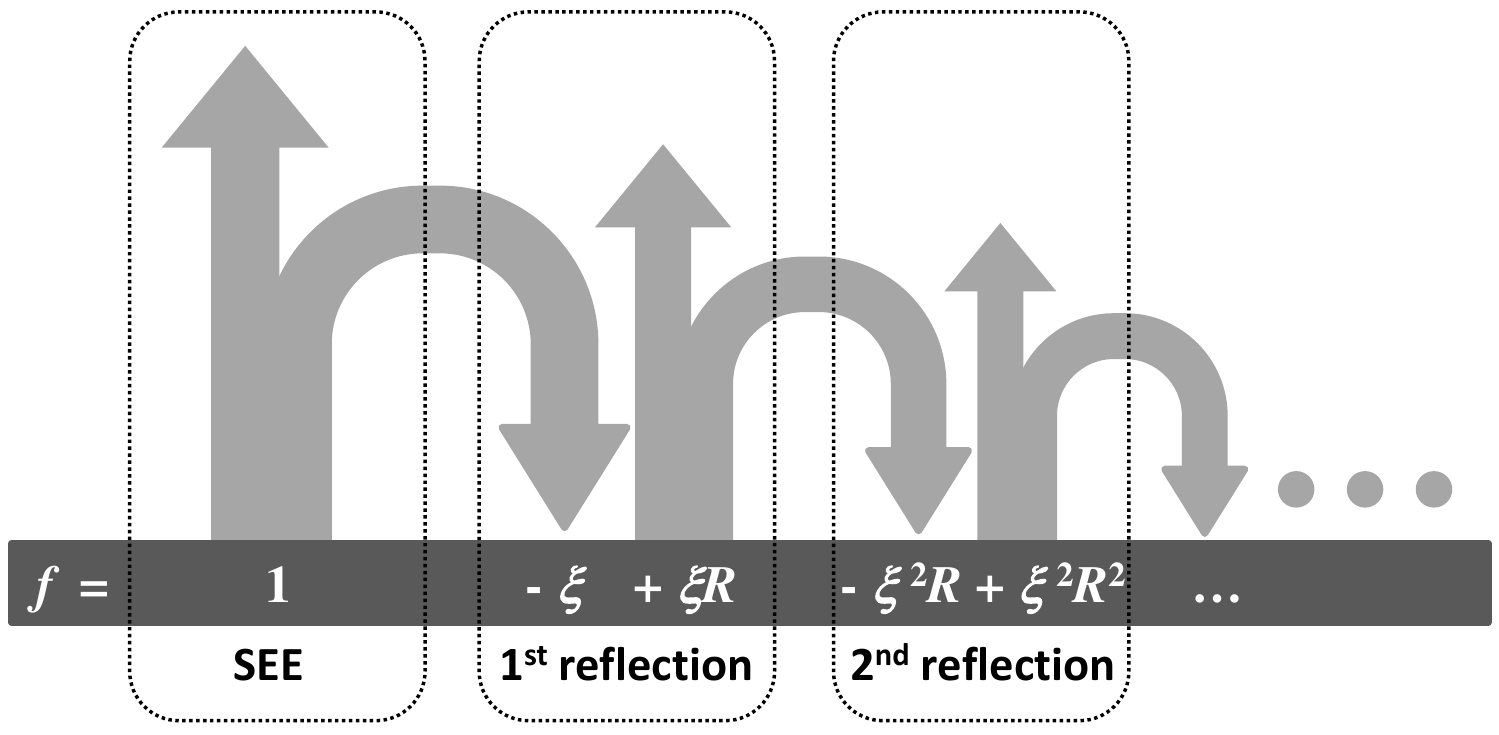}
\centering
\caption{Schematic representation of the SEE process with multiple reflections.}
\label{fig:Fig1}
\end{figure}

The first step in obtaining an accurate analytical expression for the relative SEY $f$ is the analysis of the reflection process, schematically shown in Fig. \ref{fig:Fig1}. Without a magnetic field, the relative SEY is $f=1$. With a magnetic field, a certain fraction of the secondary electrons $\xi$ returns to the surface. This fraction is more important as the angle of the magnetic field $\theta_B$ increases. From the returned fraction $\xi$, a sub-fraction $\xi R$ is reflected back into the simulation space. Thus, after the first reflection, the relative SEY loses the fraction $\xi(1-R)$. The reflected sub-fraction $\xi R$ will experience the same cycle. After $n$ successive reflections on the surface, the relative SEY can be written as:
\begin{equation}
f=1-\xi(1-R)-\xi^2 R(1-R)-...-\xi^n R^{n-1} (1-R)
\label{eq:4},
\end{equation}
which is a power series having the sum:
\begin{equation}
f=1-\xi(1-R)\frac{1-(\xi R)^n}{1-\xi R}
\label{eq:5}.
\end{equation}
Since both $\xi$ and $R$ are smaller than 1, the term $(\xi R)^n$ tends to zero for an infinite number of reflections $(n\rightarrow\infty)$ and the relation (\ref{eq:5}) converges to:
\begin{equation}
f=\frac{1-\xi}{1-\xi R}
\label{eq:6}.
\end{equation}
 
\begin{figure}
\includegraphics[width=0.95\textwidth]{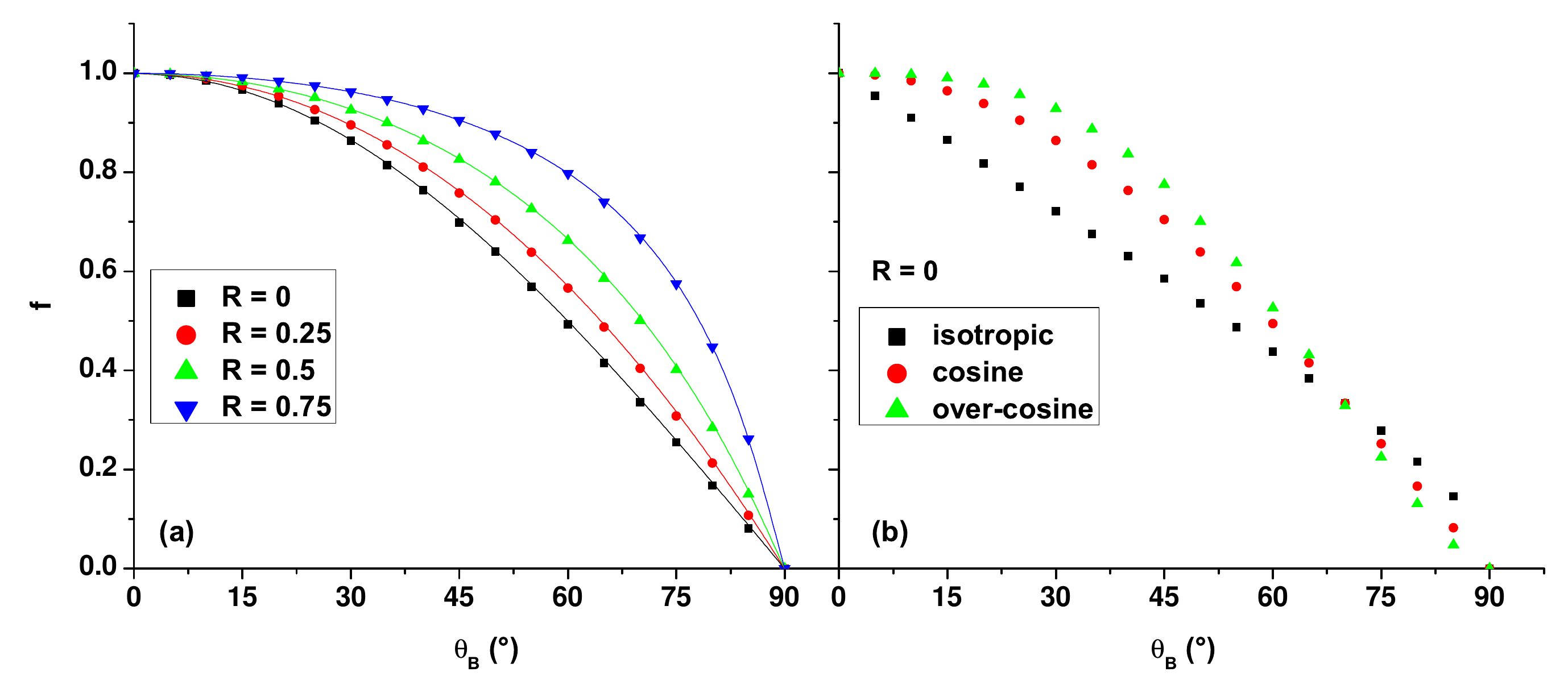}
\centering
\caption{Dependence of the relative SEY on the magnetic field angle $\theta_B$, in the absence of an electric field: (a) secondary electrons with cosine angular distribution and different values of the reflection coefficient $R$; (b) secondary electrons with different angular distributions and $R = 0$. Symbols correspond to MC simulations, full lines in panel (a) correspond to eq. (\ref{eq:9}).}
\label{fig:Fig2}
\end{figure}

Equation (\ref{eq:6}) is a generally valid formula that defines the relative SEY in the case of multiple reflections of secondary electrons on the emissive surface. It can be customized for particular cases by explicitly including the returning fraction $\xi$. In the limit case of $R=1$, the relative SEY is 1, regardless of the value of any other parameter. In the absence of an electrostatic sheath to the surface ($E=0\text{ V/m}$) and without reflection ($R=0$), MC calculations show (Fig. \ref{fig:Fig2}(a)) that the relative SEY is described by:
\begin{equation}
f|_{E=0,R=0}=\cos\theta_B
\label{eq:7}.
\end{equation}
Based on eqs. (\ref{eq:6}) and (\ref{eq:7}), the fraction of secondary electrons that return to the surface due to the magnetic field is:
\begin{equation}
\xi|_{E=0}=1-\cos\theta_B
\label{eq:8}.
\end{equation}
Introducing (\ref{eq:8}) in (\ref{eq:6}) we obtain:
\begin{equation}
f|_{E=0}=\frac{\cos\theta_B}{1-R(1-\cos\theta_B)}
\label{eq:9}.
\end{equation}
Equation (\ref{eq:9}) is the analytical formula of the relative SEY in an oblique magnetic field, without an electric field to the surface, for a flat surface that emits and reflects electrons with a cosine angular distribution. It also applies when the electron Larmor radius is much larger than the sheath thickness (usually at lower magnetic fields) and the effect of the electric field becomes negligible. Fig. \ref{fig:Fig2}(a) shows a perfect agreement between the analytical expression (\ref{eq:9}) and the MC calculations for all possible angles and different electron reflection coefficients. For $R = 0$, the reported results are in perfect agreement with previous findings \cite{Mizo95,Tskh00,Igit01,Sato94}, the relative SEY exhibiting the same decrease with $\theta_B$. However, none of the previous works associated the results with the analytical formula (\ref{eq:7}). The inclination of the magnetic field reduces $f$ from 1 ($\theta_B=0^{\circ}$) to 0 ($\theta_B=90^{\circ}$). When $\theta_B=0^{\circ}$, the Lorentz force is parallel to the surface, non of the secondary electrons being returned to the surface. The magnetic field does not influence the secondary electron emission and $f = 1$. When $\theta_B=90^{\circ}$, the Lorentz force acts in a plane perpendicular to the surface, all secondary electrons being returned to the surface, resulting in $f = 0$. At intermediate $\theta_B$ values, the number of returned electrons increases with the magnetic field tilt.  

The dependence of the relative SEY on $R$ (Fig. \ref{fig:Fig2}(a)) has not been investigated in the works cited for $R = 0$. It is compared with the results reported for magnetron discharges \cite{Buyl04,Cost05}, showing the same trend: the relative SEY increases with the increase of $R$. For higher reflection coefficients, more of the returned secondary electrons have the chance to be reflected back into plasma/vacuum, increasing thus the value of $f$. In the present study, the magnetic field has straight lines, having one end to the surface and one end to infinity (open lines). In magnetron discharges, most of the magnetic field lines are curved, having both ends to the surface (closed lines). Such configuration changes the interaction of secondary electrons with the surface. The interaction is non-local in magnetrons because an electron that escapes from the surface at one end of the magnetic field line can return to the surface at the other end. The comparison with magnetron discharge results should remain at the trend stage, unless the collision mean free path is shorter than the magnetic field lines.

Simulations show that, in the absence of an electric field, the relative SEY does not depend on the magnetic flux density $B$ or the EDF of the secondary electrons, in accordance with the results reported in \cite{Mizo95}. However, the relative SEY depends on the angular distribution of the secondary electrons, hence $f|_{E=0}=f(\theta_B,R,g_\Omega)$. The current results are in disagreement with the statements made in \cite{Tskh00,Sato94}, according to which the relative SEY has a weak dependence on the angular distribution. The dependence is quite important since the angular distribution sets the emission angle of each secondary electron. The emission angle determines the ratio of the two parameters that define the helical trajectory: the Larmor radius and the pitch. The mentioned ratio coupled with the magnetic field tilt is responsible for the return of the electron to the surface.

Figure \ref{fig:Fig2}(b) shows the relative SEY calculated for three angular distributions: isotropic, cosine and over-cosine ($g_\Omega(\theta)\sim\cos^{2}{\theta}$). The reflection coefficient has been set to zero. Thus, the results reflect only the influence of the angular distribution on the relative SEY. All distributions have been generated according to ref. \cite{Cost20}. With respect to the isotropic distribution, cosine-type distributions exhibit an enhanced secondary electron emission in quasi-perpendicular direction to the surface \cite{Cost20}. For cosine-type distributions and low values of $\theta_B$, more electrons have a higher velocity component parallel to the magnetic field line. This results in a higher pitch of the helical trajectory and a faster electron drift from the surface. Thus, more electrons can escape from the surface and the relative SEY becomes higher with respect to the isotropic distribution. With the increase of $\theta_B$, the velocity component parallel to the magnetic field line decreases for most of the electrons emitted with a cosine-type distribution and more of them return to the surface. Thus, for $\theta_B$ larger than $70^{\circ}$, the relative SEY is higher for the isotropic distribution than for cosine-type distributions. The effect of an over-cosine distribution on the relative SEY is stronger than that of the cosine distribution (Fig. \ref{fig:Fig2}(b)).

Even if the EDF of the secondary electrons does not explicitly appear in (\ref{eq:9}), the reflection coefficient $R$ might depend on the energy of the incident electrons \cite{Izaw95,Zimm04,Cimi04}. The secondary electrons become incident electrons when they are returned to the surface by the magnetic field. So, there might be an indirect dependence of $f|_{E=0}$ on the EDF of the secondary electrons. This aspect was not analyzed due to the large dispersion of values reported in the literature on the reflection coefficient $R$. Not only does $R$ depend on the surface material \cite{Caza12,Pati15,Bird91,Demi15} and the chemical state of the surface \cite{Cimi04,McRa76}, but even for the same material (e.g. Cu) the reported results are scattered\cite{Izaw95,Cimi04,McRa76,Gimp43}. Consequently, each case with variable $R$ should be treated separately.

In a previous work, a simple analytical formula of the relative SEY has been obtained from a fluid model \cite{Cost05}. The result is based on the difference between the secondary electron flux with and without a magnetic field. In a low-pressure approximation (the collision frequency is much lower than the cyclotron frequency), without electron reflection, the formula proposed in \cite{Cost05} is reduced to $f=(\frac{B_n}{B})^2=\cos^2\theta_B$. $B_n$ is the component of the magnetic field $B$ normal to the surface. The difference with respect to eq. (\ref{eq:7}) suggests an analogy with the difference between the classical cross-field transport, which is proportional to $B^{-2}$ and the empirical cross-field transport, which is proportional to $B^{-1}$ (the so-called Bohm diffusion). At this point, the analogy is just an assumption that deserves further investigation. Regarding the electron reflection, the analytical formula proposed in \cite{Cost05} considers a single reflection, obtaining different results with respect to this study.
 
\begin{figure}
\includegraphics[width=0.95\textwidth]{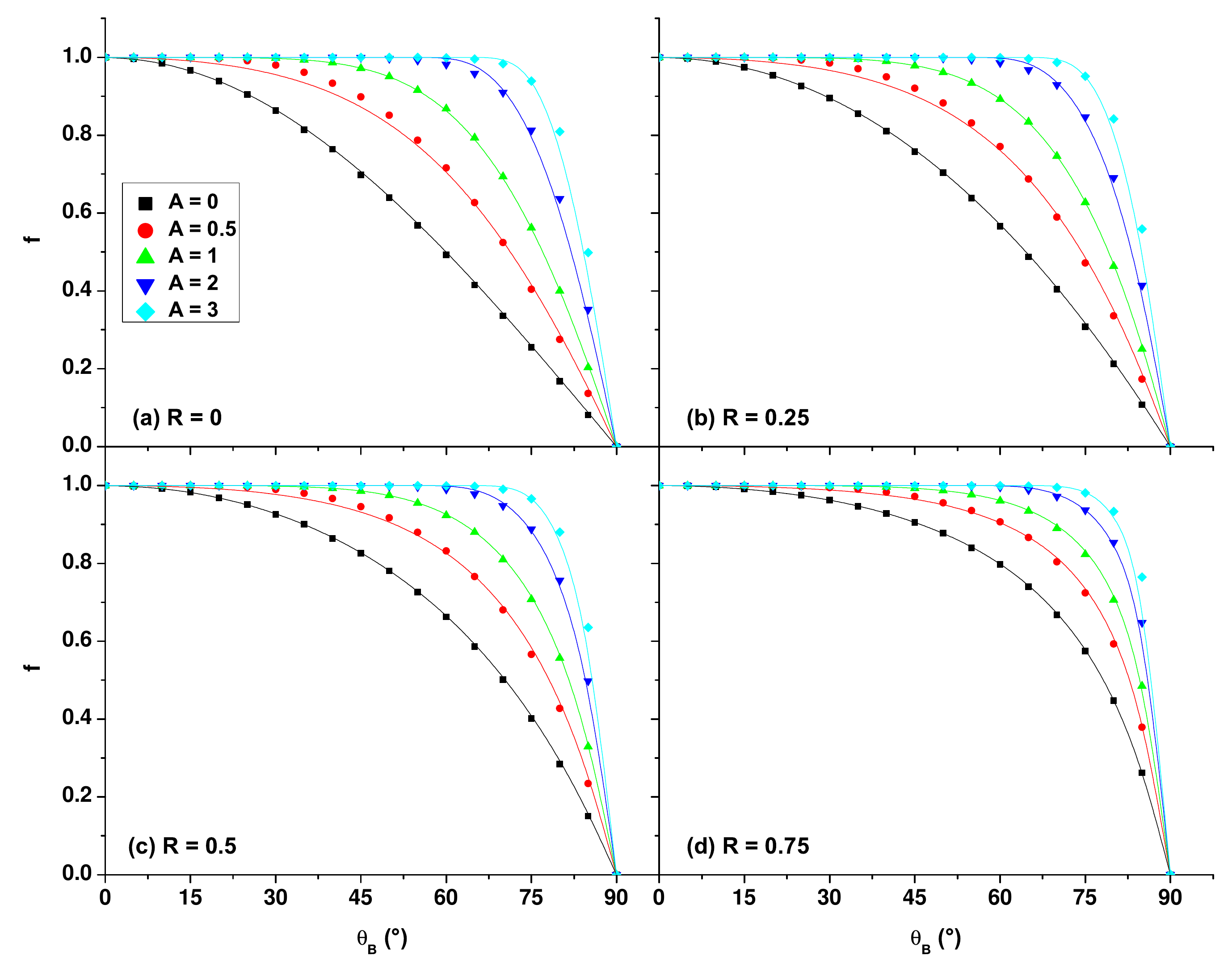}
\centering
\caption{Dependence of the relative SEY on the magnetic field angle $\theta_B$, for different values of the reflection coefficient $R$ and different values of the parameter $A$. Symbols correspond to MC simulations, full lines correspond to eq. (\ref{eq:13}). The legend in panel (a) is valid for all panels.}
\label{fig:Fig3}
\end{figure}

In the presence of an electrostatic sheath to the surface ($E\neq0\text{ V/m}$), MC calculations show that the relative SEY depends on more parameters than in eq. (\ref{eq:9}), namely $f=f(\theta_B,R,E,B,\epsilon_S,g_\Omega)$. All further calculations are made for a cosine angular distribution, so $g_\Omega$ is fixed. Analysing the emission angles of the recaptured electrons shows that the presence of the electric field is equivalent to a reduction of the magnetic field inclination with respect to the surface normal. The simulation results show that the reduced angle $\theta_{BE}$ can be expressed as:
\begin{equation}
\theta_{BE}=\theta_B(1-A\cos\theta_B)
\label{eq:10},
\end{equation}
where
\begin{equation}
A=\frac{2E}{Bv_S}
\label{eq:11}
\end{equation}
and $v_S$ is the most probable speed of the secondary electrons:
\begin{equation}
v_S=\sqrt{\frac{2e\epsilon_S}{m_e}}
\label{eq:12}.
\end{equation}
The fraction $\xi$ that returns to the surface due to the combined action of $E$ and $B$ is obtained by replacing the angle $\theta_B$ with $\theta_{BE}$ in (\ref{eq:8}). Introducing $\xi$ in (\ref{eq:6}), the simplified form of $f$ is written as:
\begin{equation}
f=\frac{\cos\theta_{BE}}{1-R(1-\cos\theta_{BE})}
\label{eq:13},
\end{equation}
with $\theta_{BE}$ given by (\ref{eq:10}). Including (\ref{eq:10})-(\ref{eq:12}) in (\ref{eq:13}), the simple dependence $f=f(\theta_{BE},R)$ turns into the more general $f=f(\theta_B,R,E,B,\epsilon_S )$. The analytical formula (\ref{eq:13}) describes the SEE process under the combined action of $E$ and $B$, for a flat surface that emits and reflects electrons with a cosine angular distribution. It provides a straightforward solution for the calculation of the relative SEY. Equation (\ref{eq:9}) is a particular case of (\ref{eq:13}) for $E=0\text{ V/m}$. Simulation results with different combinations of $E$, $B$ and $\epsilon_S$ reveal that $f$ depends only on the value of $A$, regardless of the individual values of the three parameters. Therefore, the relative SEY can be expressed as $f=f(\theta_B,R,A)$. The consistency of formula (\ref{eq:13}) is validated in Fig. \ref{fig:Fig3} by comparison with the results of the MC simulation. The relative SEY is plotted for different values of $A$ and different reflection coefficients. 

A very good agreement is found between the MC calculations and the analytical solution (\ref{eq:13}). For very small values of the relative SEY ($\theta_B$ angles close to $90^{\circ}$), the difference between numerical and analytical results may reach $10\%$. Excepting this irrelevant case, the largest deviations are below $3\%$ and they correspond to small values of $R$, $A$, and $\theta_B$ (e.g. $R=0$, $A=0.5$, and $\theta_B$ around $30^{\circ}$ in Fig. \ref{fig:Fig3}(a)). The relative SEY increases with $A$, while the general dependence on $\theta_B$ and $R$ remains as discussed for $E=0\text{ V/m}$. Figure \ref{fig:Fig3}(d) shows that a higher value of $R$ reduces the influence of $A$ on the relative SEY. Also, a higher value of $A$ reduces the influence of $R$ (see Figs. \ref{fig:Fig3}(a)-(d) for $A=3$). Individual influences of $E$, $B$ and $\epsilon_S$ on $f$ are reflected in the dependence of $f$ on $A$. They were also discussed in \cite{Mizo95}, suggesting the dependence of $f$ on the ratio
\begin{equation}
\frac{E}{B\sqrt{\epsilon_S}}
\label{eq:14},
\end{equation}
over the parameter range in which $f$ changes rapidly. As shown in Fig. \ref{fig:Fig3}, the rapid change of $f$ is characteristic to large magnetic field angles $\theta_B$. The fraction (\ref{eq:14}), which is included in $A$, was inappropriately associated with the ratio of the $\textbf{E}\times\textbf{B}$ drift speed to the emission speed of the secondary electrons \cite{Mizo95}. In fact, the variation of $f$ is more complicated than (\ref{eq:14}). 
From Fig. \ref{fig:Fig3}, it can be observed that the curves corresponding to different $A$ values can be described by the curve corresponding to $A=0$ (eq. (\ref{eq:9})) by a translation of the magnetic field angle from $\theta_B$ to a smaller angle $\theta_{BE}$. The angle translation is proportional to $A$ and it also depends on $\theta_B$. It is smaller for $\theta_B$ close to $0^{\circ}$ and $90^{\circ}$ and larger for intermediate $\theta_B$ values. Analytically, the angle translation has been found to be expressed by eq. (\ref{eq:10}). Thus, $f$ does not depend only on $A$ but on the product $A\cos\theta_B$. This indicates that the parameter that counts for the increase of $f$ in the presence of an electric field is the electric field component parallel to the magnetic field $E_{||}=E\cos\theta_B$ and not the $\textbf{E}\times\textbf{B}$ drift velocity. The drift velocity causes electrons to move along the surface, while $E_{||}$ is responsible for the acceleration of electrons along the magnetic field lines \cite{Nish96}. The two fields $E$ and $B$ act opposite. A higher magnetic flux density enforces a smaller gyration radius and a shorter cyclotron period for the secondary electrons. The shorter the cyclotron period, the more likely the electron is to return to the surface, which is reflected in a reduction of the relative SEY. A higher electric field, i.e. a higher $E_{||}$, increases the pitch of the helical trajectory, allowing secondary electrons to move away from the surface even in a short cyclotron period. As a result, the relative SEY increases with $E$. The dependence of $f$ on $\epsilon_S$ reflects the dependence of $f$ on the EDF of the secondary electrons. A higher electron emission energy reduces the relative SEY \cite{Mizo95,Nish96,Tskh00}. A higher velocity component along $B$, directed to the surface, allows secondary electrons to return to the surface even in the presence of a repelling electric field. Thus, the increase of $\epsilon_S$ diminishes the effect of the electric field.

Figure \ref{fig:Fig3} also shows that the curves corresponding to $A=0$ (no electric field to the surface) set the minimum values of the relative SEY imposed by a tilted magnetic field. These curves, described by eq. (\ref{eq:9}), have already been discussed on Fig. \ref{fig:Fig2}. For all possible combinations of $E$, $B$ and $\epsilon_S$, the value of the relative SEY is between the minimum curve $f|_{E=0}$ and the maximum curve $f=1$. Results similar to those plotted in Fig. \ref{fig:Fig3}(a) have been reported in \cite{Mizo95,Tskh00}, but for mono-energetic secondary electrons. The results of the current MC simulations become identical to those in \cite{Mizo95,Tskh00} if mono-energetic secondary electrons are implemented in the numerical code.

\begin{figure}
\includegraphics[width=8.6cm]{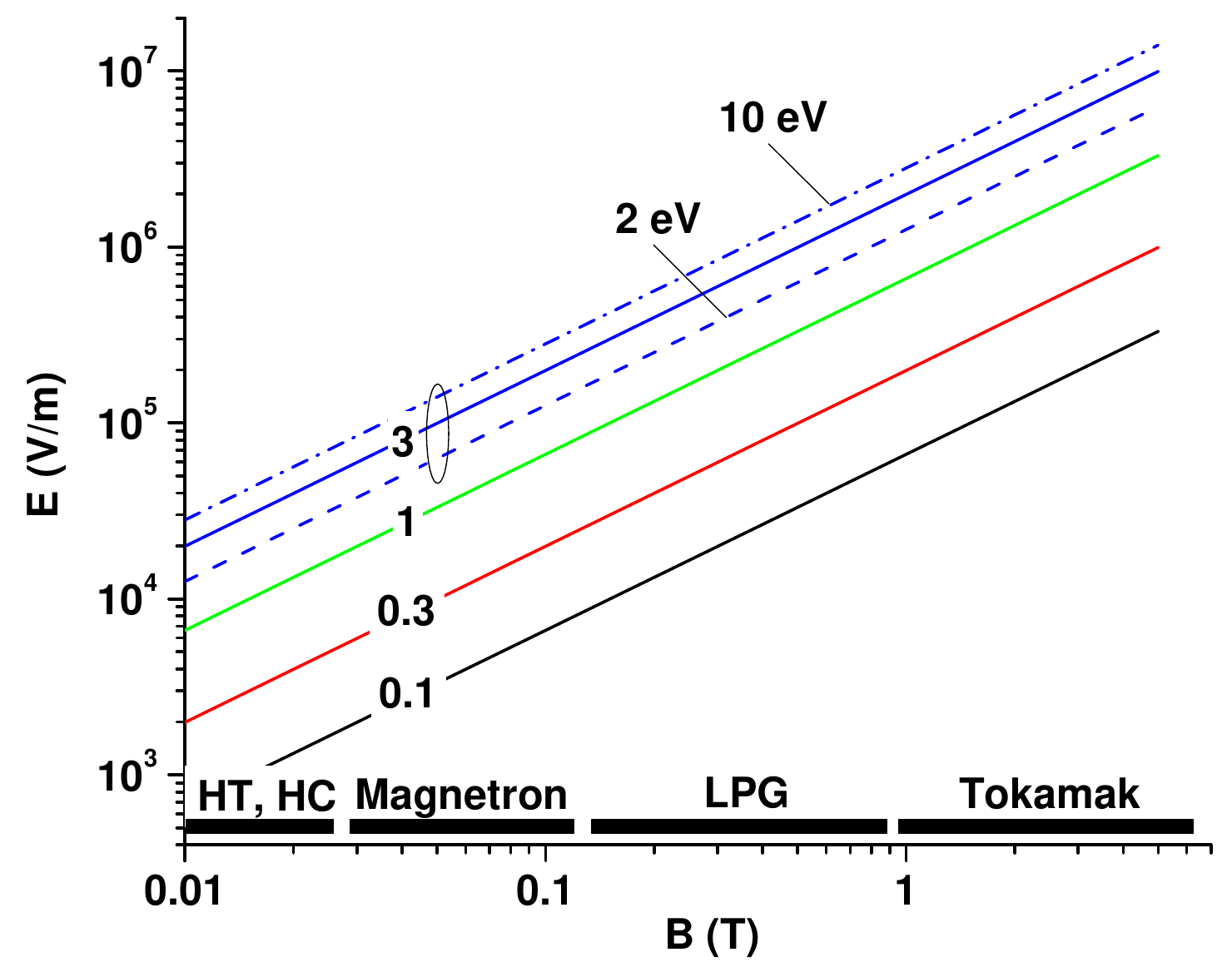}
\centering
\caption{Different electric and magnetic fields combined in (\ref{eq:11}) to get specific values of the parameter $A$ ($\epsilon_S =5\text{ eV}$ except where clearly indicated). Typical magnetic field values for Hall thrusters (HT), hollow cathode discharges (HC), magnetron sputtering devices, linear plasma generators (LPG), and tokamaks are indicated.}
\label{fig:Fig4}
\end{figure}

Possible combinations of $E$ and $B$ fields corresponding to a specific value of $A$ are shown in Fig. \ref{fig:Fig4}, for the energy $\epsilon_S=5\text{ eV}$ of the secondary electrons. The influence of $\epsilon_S$ is illustrated by plotting the curve $A=3$ for two more values of $\epsilon_S$ (2 and 10 eV). If $\epsilon_S$ increases by a factor of 2, the electric field has to increase twice in order to preserve the value of $A$. In other words, for a fixed electric field, an increase of $\epsilon_S$ by a factor of 2 results in reducing $A$ by half. In such a case, the relative SEY will decrease, as shown in Fig. \ref{fig:Fig3}. Typical magnetic field values are indicated in Fig. \ref{fig:Fig4} for different applications: $0.01-0.03\text{ T}$ for Hall thrusters (HT) and hollow cathode discharges (HC), $0.03-0.1\text{ T}$ for magnetron sputtering devices, and $1-5\text{ T}$ for tokamaks. The magnetic field range between magnetrons and tokamaks, $0.1-1\text{ T}$, is covered by magnetically confined linear plasma generators (LPG) \cite{Cost15,Kret15}. Such devices are specially designed to investigate plasma-surface interactions which are relevant for edge regions of fusion reactors \cite{Cost16}. Whilst the magnetic field is externally imposed, the electric field in front of the surface depends on plasma parameters. Customized electric field values can be obtained if plasma density, electron temperature and surface bias with respect to plasma are known. For an electric field of $10^5\text{ V/m}$ (a relatively common value for the mentioned applications) and $\epsilon_S =5\text{ eV}$, the values of $A$ are: $A>3$ for HT, $A\approx3$ for magnetrons, $A\approx1$ for linear devices and $A\approx0.1$ for tokamaks. The relative SEYs plotted in Fig. \ref{fig:Fig3} correspond very well to this example. Thus, for tokamaks, an electric field of $10^5\text{ V/m}$ is sufficiently weak not to influence the relative SEY. However, it is very strong for Hall thrusters and magnetrons, completely cancelling the influence of the magnetic field for tilt angles below $70-75^{\circ}$. The latter statement is valid only if the sheath thickness is twice as large as the Larmor radius of the secondary electrons. Otherwise, a thinner sheath results in a lower relative SEY. This is because the electric field influences only a part of the electron trajectory. A sheath contraction is equivalent to a reduction of the electric field in large sheaths. The sheath thickness is usually a few Debye lengths. The Debye length is approximately $7.4\sqrt{\frac{Te}{n_0}}$, where $T_e$ is the temperature of plasma electrons, expressed in energy units eV, and $n_0$ is the plasma density in cm$^{-3}$. The Larmor radius can be roughly estimated as $3.4\times10^{-6}\frac{\sqrt{\epsilon_S}}{B}$, with $\epsilon_S$ in eV and $B$ in T.

Equation (\ref{eq:10}) has physical meaning if $\theta_{BE}$ is positive. When $\theta_{BE}$ becomes negative, or
\begin{equation}
A\cos\theta_B>1
\label{eq:15},
\end{equation}
none of the secondary electrons return to the surface and the relative SEY is equal to 1. It occurs when the electric field component $E_{||}$ is strong enough to move all secondary electrons away from the surface, regardless of their emission angle. In such a case, the effect of the magnetic field on the SEE is completely suppressed. The inequality (\ref{eq:15}) is fulfilled for $A$ larger than 1 and $\cos\theta_B >1/A$. For example, in Fig. \ref{fig:Fig3}(b), $f=1$ for $A=2$ and $\theta_B <60^{\circ}$ or $A=3$ and $\theta_B <70.5^{\circ}$. Figure \ref{fig:Fig4} shows that $A$ larger than 1 is obtained for electric fields larger than $\sim10^4\text{ V/m}$ in HT and $\sim10^6\text{ V/m}$ in tokamaks. 

\begin{figure}
\includegraphics[width=8.6cm]{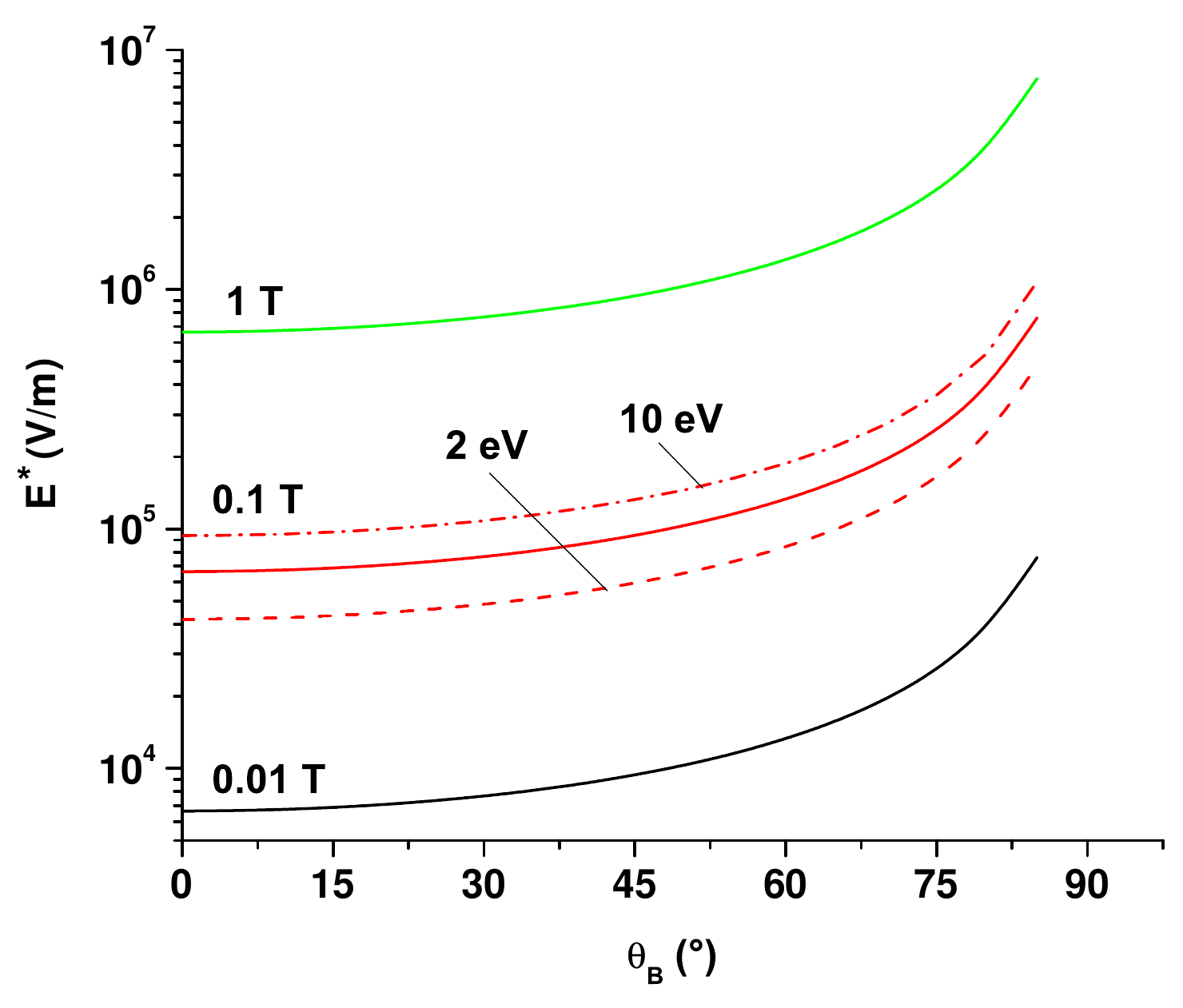}
\centering
\caption{Electric field limit $E^{*}$ calculated from (\ref{eq:16}) as a function of the magnetic field angle $\theta_B$, for different magnetic flux densities ($\epsilon_S =5\text{ eV}$ except where clearly indicated).}
\label{fig:Fig5}
\end{figure}

Including (\ref{eq:11})-(\ref{eq:12}) in (\ref{eq:15}), an electric field limit $E^*$ can be calculated:
\begin{equation}
E^*=\frac{B}{2\cos\theta_B}\sqrt{\frac{2e\epsilon_S}{m_e}}
\label{eq:16}.
\end{equation} 
Above $E^*$, the effect of the magnetic field on the SEE is suppressed. The value of $E^*$ is plotted in Fig. \ref{fig:Fig5} as a function of $\theta_B$ angle, for different magnetic flux densities (the same orders of magnitude as for HT, magnetrons and tokamaks) and for $\epsilon_S =5\text{ eV}$. As in Fig. \ref{fig:Fig4}, the influence of $\epsilon_S$ is illustrated by plotting the curve $B=0.1\text{ T}$ for two more values of $\epsilon_S$ (2 and 10 eV). At high magnetic field angles $\theta_B >80^{\circ}$, the electric field limit $E^*$ increases by an order of magnitude. $E^*$ is not defined when the magnetic field is parallel to the surface ($\theta_B =90^{\circ}$) but, in this case, the relative SEY is zero for all conditions. Reminder: results were obtained assuming non-collisional electron trajectories (very low pressure) and an electrostatic sheath thickness larger than twice the Larmor radius. 

\section*{Conclusion}

The relative secondary electron emission yield in an oblique magnetic field can be calculated with formula (\ref{eq:13}) which has been derived based on the results of Monte Carlo simulations. The analytical formula is valid for a flat surface, open magnetic field lines, a constant electric field in the sheath and a cosine angular distribution of the secondary electrons. The magnetic flux density and the emission energy of the secondary electrons contribute to the reduction of the relative SEY. Electron reflection coefficient on the surface acts the opposite. The magnetic field tilt with respect to the surface normal has a major influence on the effective emission. An electric field reduces the magnetic field effect, equivalent to a reduction of the magnetic field tilt. Without electric field, the relative SEY depends on the magnetic field angle, reflection coefficient, and the angular distribution of the secondary electrons. Formula (\ref{eq:13}) is a reliable tool for studying the implications of an effective SEE in magnetically assisted devices (tokamaks, magnetrons, Hall thrusters), in scanning electron microscopy, in electron cloud mitigation etc, helping the design of such applications. It also provides realistic input for simulations of already mentioned applications, especially for 0D and 1D codes that are not able to describe the effective SEE process. A very good agreement has been found between the MC simulations and the analytical formula. Further studies should aim investigating the influence of the background pressure on the relative SEY. A variable electric field within the electrostatic sheath or a self-consistent model of the sheath may also be considered.

\bibliography{Costin.bib}

\section*{Acknowledgements}

This work has been carried out within the framework of the EUROFusion Consortium and has received funding from the Euratom research and training programme 2014-2018 and 2019-2020 under grant agreement No 633053. The views and opinions expressed herein do not necessarily reflect those of the European Commission.

\section*{Additional information}

\subsection*{Author Contribution statement}

The author confirms sole responsibility for the entire work and manuscript.

\subsection*{Competing interests}

The author declares no competing interests.

\end{document}